\documentclass[12pt,notitlepage]{article}
\usepackage{nopageno}
\usepackage[top=1in,left=1.25in,right=1.25in,bottom=1in]{geometry}
\usepackage[scriptsize,aboveskip=0pt,belowskip=0pt]{caption}
\usepackage[T1]{fontenc}
\usepackage{amsthm}
\makeatletter
\def\thm@space@setup{\thm@preskip=0pt
\thm@postskip=0pt}
\makeatother

\usepackage{graphicx}
\usepackage{amssymb}
\usepackage{url}
\usepackage{amsmath}
\usepackage{algorithm}
\usepackage[noend]{algpseudocode}
\usepackage{float,color,subfigure}
\usepackage{txfonts}
\usepackage[T1]{fontenc}
\usepackage{setspace}
\usepackage[small,compact]{titlesec}
\titlespacing*{\section}{0pt}{1pt}{1pt}
\usepackage[affil-it]{authblk}

\usepackage{abstract}
\usepackage{titling}
\usepackage[
 natbib=true,
 sortcites=true,
 style=numeric,
 maxnames=2,
 firstinits=true,
 doi=false,isbn=false,url=false,
 backend=bibtex]{biblatex}

\renewcommand*{\intitlepunct}{\addspace}
\renewbibmacro{in:}{%
  \ifentrytype{article}{}{\printtext{\bibstring{in}\intitlepunct}}}
\newcommand{\Str}{\operatorname{\sf S}}
\newcommand{\rank}{\operatorname{\sf rank}}
\newcommand{\select}{\operatorname{\sf select}}
\newcommand{\access}{\operatorname{\sf access}}

\newtheorem{theorem}{Theorem}[section]

\newcommand{\subsec}[1]{\smallskip \noindent {\bf #1}. }

 \usepackage[hang,flushmargin]{footmisc} 

\usepackage{etoolbox} % modify commands and many other things
\makeatletter
\patchcmd\maketitle{\@makefntext}{\@@@ddt}{}{}
\patchcmd\maketitle{\rlap}{\mbox}{}{}
\makeatother

\posttitle{\par\end{center}\vspace{-8pt}} \title{Parallel Wavelet Tree
  Construction
\thanks{\scriptsize This is a longer version of the paper that appears in the \emph{Proceedings of the IEEE Data Compression Conference, 2015}. The levelWT algorithm has been slightly simplified.} 
\thanks{\scriptsize 
We thank Simon Gog and Matthias
    Petri for discussions, helping with the
    code in~\cite{Gog2014}, and providing the \emph{trec8} data set,
    Leo Ferres and Jose Fuentes-Sepulveda for discussions and providing the
    code from~\cite{Fuentes14}, and Guy Blelloch for discussions.}} 

\author{\vspace{-5pt} Julian Shun\thanks{\scriptsize Supported by the National
Science Foundation under grant number CCF-1314590.
}\authorcr \vspace{-5pt}{\small \emph{Carnegie Mellon University}} \authorcr {\small \emph{jshun@cs.cmu.edu}}}%\\Carnegie Mellon University\\jshun@cs.cmu.edu}

\date{}
\long\def\symbolfootnote[#1]#2{\begingroup%
\def\thefootnote{\fnsymbol{footnote}}\footnote[#1]{#2}\endgroup} 

\usepackage{txfonts}                            % otherwise, Adobe Times and Math Fonts
\usepackage{txgreeks}

\usepackage{microtype}

\begin{document}

\maketitle

%\vspace{-9.9mm}
\renewcommand{\abstractname}{}    % clear the title
\renewcommand{\absnamepos}{empty} % originally center

\vspace{-32pt}
\begin{abstract}
%% Wavelet trees have received significant attention due to their
%% applications in compressed data structures. 
We present
parallel algorithms for wavelet tree
construction with polylogarithmic depth, 
%% Our first algorithm is based on constructing
%% the wavelet tree level-by-level. For an input size of $n$ and alphabet
%% size $\sigma$, our first algorithm uses $O(n\log\sigma)$ work, which
%% matches that of the standard sequential algorithm, and $O(\log
%% n\log\sigma)$ depth.  Our second algorithm constructs all levels of the
%% tree in parallel, and requires $O(W_{sort}\sigma)$ work and
%% $O(S_{sort}+\log n)$ depth where $W_{sort}$ and $S_{sort}$ are the work
%% and depth, respectively, of the stable parallel integer sorting routine
%% used. This is either $O(n\log\log n\log\sigma)$ work and $O(\log n)$
%% depth for all alphabet sizes or $O(n\log\sigma)$ work and
%% $O(\sigma^{\epsilon}+\log n)$ depth ($0 < \epsilon < 1$). For alphabets
%% of polylogarithmic size, this is $O(n\log\sigma)$ work and $O(\log n)$
%% depth.  We then discuss a variant of the second algorithm that requires
%% $O(n\log\sigma)$ work and $O(\log n\log\sigma)$ depth.  
improving upon the linear depth of the recent parallel algorithms by
Fuentes-Sepulveda et al. We experimentally show on a 40-core
machine with two-way hyper-threading that we outperform the existing
parallel algorithms by 1.3--5.6x and achieve up to 27x speedup over
the sequential algorithm on a variety of real-world and artificial
inputs. Our algorithms show good scalability with increasing thread count,
input size and alphabet size.  We also discuss extensions
to variants of the standard wavelet tree.
%% We
%% also show that structures supporting constant time rank/select
%% queries, which are used in wavelet trees, can be constructed in linear
%% work and logarithmic depth.
\end{abstract}

\vspace{-8pt}
\section{Introduction}
The \emph{wavelet tree} was first described by Grossi et
al.~\cite{Grossi2003}, where it was used in compressed suffix
arrays. It is a space-efficient data structure that supports access,
rank and select queries on a sequence in $O(\log\sigma)$ work, where
$\sigma$ is the alphabet size of the sequence.  Since its initial use,
wavelet trees have found many other applications, for example in
compressed representations of sequences, permutations, grids, graphs,
self-indexes based on the Burrows-Wheeler transform~\cite{Burrows94},
images, two-dimensional range queries~\cite{MakinenN07}, among many
others (see~\cite{Navarro2012,Makris12} for surveys of applications).
%% Makinen and
%% Navarro~\cite{MakinenN07} observed that wavelet trees can also be used
%% for two-dimensional range queries.
%% , and relates it to a similar
%% structure described by Chazelle~\cite{Chazelle1988}.
While applications of wavelet trees have attracted significant
attention, wavelet tree construction has not been widely studied. This is not
surprising, as the standard sequential algorithm for wavelet tree
construction
%, which requires $O(n\log\sigma)$ work for a sequence of
%length $n$, 
is very straightforward.
The algorithm requires $O(n\log\sigma)$
work for a sequence of length $n$.
However, constructing the wavelet tree of large sequences (with
large alphabets) can be time-consuming, and hence parallelizing the
construction is important. A step in this direction was taken recently
by Fuentes-Sepulveda et al.~\cite{Fuentes14}, who describe parallel
algorithms for constructing wavelet trees that require $O(n)$ depth
(number of parallel time steps).

In this paper, we describe parallel algorithms for wavelet tree
construction that exhibit much more parallelism (in particular, polylogarithmic depth). 
%% , in contrast to the
%% algorithms of Fuentes-Sepulveda et al. that require $O(n)$ depth. 
We
first describe an algorithm that constructs the tree level-by-level,
and requires $O(n\log\sigma)$ work and $O(\log n\log\sigma)$ depth. We
then describe a second algorithm that requires
$O(W_{sort}(n)\log\sigma)$ work and $O(D_{sort}(n)+\log n)$ depth, where
$W_{sort}(n)$ and $D_{sort}(n)$ are the work and depth, respectively,
of the parallel stable integer sorting routine used in the
algorithm. Using a linear-work integer sort~\cite{RR89}, we
obtain a work bound of $O(\frac{n}\epsilon\log\sigma)$ and depth bound of
$O(\frac{1}\epsilon(\sigma^{\epsilon}+\log n))$ for some constant $0 < \epsilon < 1$,
which is sub-linear.  For alphabets of polylogarithmic size, this
gives an algorithm with $O(\log n)$ depth. Using a super-linear work
integer sort~\cite{Raman1990,Bhatt1991}, we can obtain a work bound of
$O(n\log\log n\log\sigma)$ and depth bound of $O(\log n)$ for all
alphabets.
%% We describe a variant of this algorithm that requires
%% $O(n\log\sigma)$ work and $O(\log n\max(1,\log\sigma/\log\log n))$
%% depth for all alphabets.  
In addition to having good theoretical bounds, our algorithms are also
efficient in practice. We implement our algorithms using Cilk Plus and
show experiments on a 40-core shared-memory machine (with two-way
hyper-threading) indicating that they outperform the existing parallel
algorithms for wavelet tree construction by 1.3--5.6x and achieve up
to 27x speedup over the sequential algorithm. We show that our
implementations scale well with increasing thread count, input size and
alphabet size.  We then discuss the parallel construction of
rank/select structures on binary sequences, which are an essential
component to wavelet trees.  Finally, we discuss how to adapt our
algorithms to variants of wavelet trees---Huffman-shaped wavelet
trees~\cite{Foschini2006}, multiary wavelet
trees~\cite{Ferragina2007}, and wavelet matrices~\cite{Claude12}.
%% In~\cite{full}, we also discuss how to
%% parallelize a very recent $O(n\log\sigma/\sqrt{\log n})$ work wavelet
%% tree algorithm by Babenko et al.~\cite{BabenkoGKS14}, and obtain a
%% parallel algorithm with $O(n\log\sigma/\sqrt{\log n})$ work and
%% $O(\log n\log\sigma)$ depth, although we have not yet investigated the
%% practicality of the algorithm.

\section{Preliminaries}\label{sec:prelims}
We state complexity bounds of algorithms in the work-depth model, where
the \emph{work} $W$ is the number of operations required
and the \emph{depth} $D$ is the number of time steps
required. Then if $P$ processors are available, using Brent's
scheduling theorem~\cite{JaJa92}, we can bound the running time by
$O(W/P + D)$. The parallelism of an algorithm is equal to
$O(W/D)$.  We allow for concurrent reading and writing in the model.
%% For sequential algorithms, the work and depth are
%% equivalent.
%% We say a parallel algorithm is \emph{work-efficient} if
%% its asymptotic work complexity matches that of the best sequential
%% algorithm.

We denote a sequence by $\Str$, where $\Str[i]$ is the $i$'th symbol
of $\Str$ and $n$ is its length. We denote an alphabet by $\Sigma =
[0,\ldots,\sigma-1]$, where $\sigma$ is the alphabet size.
%% For a sequence with alphabet $\Sigma$, $s_i \in \Sigma$ for
%% all $i$.
$\access(\Str,i)$ returns the symbol
at position $i$ of $\Str$, $\rank_c(\Str,i)$ returns the
number of times $c$ appears in $\Str$ from positions 0 to $i$, and
$\select_c(\Str,i)$ returns the position of the $i$'th
occurrence of $c$ in $S$.

A \emph{wavelet tree} is a data structure that supports access, rank
and select operations on a sequence in $O(\log\sigma)$ work (we use
$\log x$ to mean the base 2 logarithm of $x$, unless specified
otherwise).  The standard wavelet tree is a binary
tree where each node represents a range of the symbols in $\Sigma$
using a bitmap (binary sequence).  We assume $\sigma \le n$ as the
symbols can be mapped to a contiguous range otherwise.  The structure
of the wavelet tree is defined recursively as follows: The root
represents the symbols $[0,\ldots,2^{\lceil \log \sigma \rceil}-1]$.
A node $v$ which represents the symbols $[a,\ldots,b]$ stores a bitmap
which has a $0$ in position $i$ if the $i$'th symbol in the range
$[a,\ldots,b]$ is in the range $[a,\ldots,\frac{(a+b+1)}2-1]$, and $1$
otherwise. It will have a left child that represents the symbols
$[a,\ldots,\frac{(a+b+1)}2-1]$ and a right child that represents the
symbols $[\frac{(a+b+1)}2,\ldots,b]$. The recursion stops when the
size of the range is 2 or less or if a node has no symbols to
represent.  We note that the original wavelet tree description
in~\cite{Grossi2003} uses a root whose range is not necessarily a
power of 2.  However, the definition we use gives the same query
complexities and leads to a simpler description of our algorithms.

Along with the bitmaps, each node stores a \emph{succinct} rank/select
structure (whose size is sub-linear in the bitmap length) to allow for
constant work rank and select queries. The structure of a wavelet tree
requires $n\lceil \log \sigma \rceil + o(n\log\sigma)$ bits (the lower
order term is for the rank/select structures). The tree topology
(parent and child pointers) requires $O(\sigma\log n)$ bits, though
this can be reduced or removed by modifying the queries
accordingly~\cite{MakinenN07,ClaudeN08}.
The standard sequential algorithm for wavelet tree construction takes
$O(n\log\sigma)$ work. 
%% It starts from the root of the tree, and
%% constructs the tree from top to bottom.  For each node, it reads its
%% sequence to fill in its bitmap and also partitions the sequence into
%% two sub-sequences, which are passed to its two children nodes. The
%% rank/select structure per node can be constructed in work linear in
%% its bitmap length.

In this paper, we will use the basic parallel primitives, prefix sum and
filter~\cite{JaJa92}. \emph{Prefix sum} takes an array $X$ of
length $n$, an associative binary operator $\oplus$, and an identity
element $\bot$ such that $\bot \oplus x = x$ for any $x$, and returns
the array $(\bot,\bot \oplus X[0], \bot \oplus X[0] \oplus X[1],
\ldots,\bot \oplus X[0] \oplus X[1] \oplus \ldots \oplus X[n-2])$, as
well as the overall sum $\bot \oplus X[0] \oplus X[1] \oplus \ldots
\oplus X[n-1]$. We will use $\oplus$ to be the $+$ operator on
integers. \emph{Filter} takes an array $X$ of length $n$, a predicate
function $f$ and returns an array $X'$ of length $n' \le n$ containing the
elements in $x \in X$ such that $f(a)$ returns true, in the same order
that they appear in $X$. Filter can be implemented using prefix sum,
and both require $O(n)$ work and $O(\log n)$ depth~\cite{JaJa92}.

\section{Related Work}
Fuentes-Sepulveda et al.~\cite{Fuentes14} describe a parallel
algorithm for constructing a wavelet tree. They observe that for an
alphabet where the symbols are contiguous in $[0,\sigma-1]$, the node
at which a symbol $s$ is represented at level $i$ of the wavelet tree
can be computed as $s \gg \lceil \log\sigma \rceil-i$, requiring
constant work. With this observation they can compute the bitmaps of
each level independently. Each level is computed sequentially,
requiring $O(n)$ work and depth. Thus, their algorithm requires an
overall work of $O(n\log\sigma)$ and $O(n)$ depth. They describe a
second algorithm which splits the input sequence into $P$
sub-sequences, where $P$ is the number of processors available. In the
first step, the wavelet tree for each sub-sequence is computed
sequentially and independently. Then in the second step, the partial
wavelet trees are merged. The merging step is non-trivial and requires
$O(n)$ depth. Thus the algorithm again requires $O(n\log\sigma)$ work
and $O(n)$ depth. This algorithm was shown to perform better than the
first algorithm due to the high parallelism in the first step.

Multiple queries on the wavelet tree can be answered in parallel since
they do not modify the tree.  Furthermore, they can be batched to take
advantage of cache locality~\cite{Fuentes14}.

%% Since individual queries require sequential traversal of the wavelet
%% tree, they cannot be parallelized. However multiple queries can be
%% answered in parallel since they do not modify the tree. Furthermore,
%% they can be batched to take advantage of cache
%% locality~\cite{Fuentes14}.

Arroyuelo et al.~\cite{Arroyuelo2012} explore the use of wavelet trees
in distributed search engines. They do not construct the wavelet tree
for the entire text in parallel, but instead sequentially construct
the wavelet tree for parts of the text on each machine.

Tischler~\cite{Tischler2011} and Claude et al.~\cite{Claude2011}
discuss how to reduce the space usage of sequential wavelet tree
construction.  Foschini et al.~\cite{Foschini2006} describe an
improved algorithm for sequentially constructing the wavelet tree in
compressed format, requiring $O(n + \min(n,nH_h)\log\sigma)$ work,
where $H_h$ is the $h$'th order entropy of the input. 
%% This bound can
%% be better than $O(n\log\sigma)$ for compressible sequences (though
%% still $O(n\log\sigma)$ in the worst case). 
The approach only works if the object produced is the wavelet tree
compressed using run-length encoding. 
%% Parallelizing these techniques
%% is a direction for future work.  
Very recently,
%% and independently of our
%% work, 
Babenko et al.~\cite{BabenkoGKS14} and Munro et al.~\cite{Munro15}
describe sequential wavelet tree construction algorithms
that require $O(n\log\sigma/\sqrt{\log n})$ work. The algorithms pack
small integers into words, and require extensive bit manipulation. As
far as we know, there are no implementations of the algorithms
available. 
%% We are interested in designing practical (parallel)
%% implementations of the algorithms in the future.
%% We discuss how to
%% parallelize their algorithm in the full version of the paper~\cite{full}.

%% Succinct data structures supporting rank and select queries on binary
%% sequences in constant time have been widely studied (see
%% e.g.~\cite{Jacobson1988,Clark1996}).
%% They have many uses in succinct data structures, and are an essential
%% component of wavelet trees. There has also been significant research
%% on adapting these structures to
%% practice (see e.g.~\cite{ZhouAK13,GogP2013}).

%~\cite{Navarro12,ZhouAK13,OkanoharaS07,Gonzalez05,ClaudeN08,Vigna2008,Ladra2012}. 
%% Besides
%% wavelet trees, rank and select structures on larger alphabets have
%% been described by Golynski et al.~\cite{Golynski2006} and Barbay et
%% al.~\cite{Barbay10}. These structures allow queries in
%% $O(\log\log\sigma)$ time, asymptotically faster than wavelet
%% trees. However, experiments show that they are competitively only on
%% very large alphabets~\cite{ClaudeN08}. 
%% The space and query times of
%% wavelet trees with different shapes and using different compression
%% techniques were experimentally studied by Grossi et
%% al.~\cite{Grossi2011}.

%\input{notation}
\section{Parallel Wavelet Tree Construction}\label{sec:alg}
We now describe our algorithms for wavelet tree
construction. The construction requires a rank/select data structure
for binary sequences. For now we assume that such structures can be
created in linear work and logarithmic depth, and defer the discussion
to Section~\ref{sec:rank-select}.

Our first algorithm, \emph{levelWT}, constructs the wavelet tree
level-by-level. On each level, the nodes and their bitmaps are
constructed in parallel in $O(n)$ work and $O(\log n)$ depth, which
gives an overall complexity of $O(n\log\sigma)$ work and $O(\log
n\log\sigma)$ depth since there are $O(\log\sigma)$ levels in the tree.
The pseudo-code for levelWT is shown in Figure~\ref{alg:levelWT}.  We
maintain a bitmap $B$ of length $n$ shared by all nodes on each level
(Line 4). Nodes will simply store its starting point in this array along
with its bitmap length.  To keep track of which nodes need to be
constructed on each level, we maintain an array $A$ of information for
nodes to be added at the next level. Each entry in $A$ stores the
starting point (\emph{start}) in the level bitmap (also the starting point in the sequence for the level) for the node,
bitmap length (\emph{len}), node identifier (\emph{id}) and length of
the alphabet range that it represents (\emph{range}). We represent an
entry of $A$ as a 4-tuple (start, len, id, range).

\begin{figure}[!ht]\scriptsize
\vspace{-3pt}
\begin{algorithmic}[1]
\Procedure{levelWT}{$\Str$, $n$, $\sigma$}
\State Nodes = $\{\}$,\hspace{7pt} $L = \lceil \log \sigma \rceil$, \hspace{7pt} $\Str' = \Str$, \hspace{7pt} $A = \{(0,n,0,2^L)\}$,\hspace{7pt} $A' = \{\}$
\For {$l = 0$ to $L-1$} \Comment{Process level-by-level}
\State mask = $2^{L-(l+1)}$,\hspace{7pt} $B =$ bitmap of length $n$ \Comment{Mask and bitmap for this level}
%% \State $O = \{\}$ \Comment{Used to store offsets into bitmap} 
%% \ParFor {$j=0$ to $|A|-1$} \hspace{3pt} $O[j] = A[j].\mbox{len}$ 
%% \EndFor
%% \vspace{-2pt}
%% \State Perform prefix sum on $O$ to get offsets into $B$
\ParFor {$j=0$ to $|A|-1$}
\State  start = $A[j].\mbox{start}$,\hspace{7pt} len = $A[j].\mbox{len}$,\hspace{7pt} id = $A[j].\mbox{id}$,\hspace{7pt} $r = A[j].\mbox{range}$ 
\State Nodes[id].bitmap = $B+\mbox{start}$,\hspace{7pt} Nodes[id].len = len
\If {$r \le 2$} \Comment{Node has no children}
\ParFor {$i=0$ to $\mbox{len}-1$}
\If {$(\Str[\mbox{start}+i] \ \&\ \mbox{mask} \neq 0)$} 
$B[\mbox{start}+i] = 1$ 
{\bf else}
$B[\mbox{start}+i] = 0$
\EndIf
\EndFor
\vspace{-2pt}
\Else
\State $X = \{\}$ \Comment{Array used to store target positions into $\Str'$}
\ParFor {$i=0$ to $\mbox{len}-1$}
{$\{$ {\bf if} {$(\Str[\mbox{start}+i] \ \&\ \mbox{mask} = 0)$} $X[i] = 1$ {\bf else} $X[i] = 0$ $\}$}
%\EndIf
\EndFor
\vspace{-2pt}
\State Perform prefix sum on $X$ to get offsets of ``left'' characters ($X_s$ is the total sum, i.e. number of ``left'' characters)
\ParFor {$i=0$ to $\mbox{len}-1$}
\If {$(\Str[\mbox{start}+i] \ \&\ \mbox{mask} \neq 0)$} %\hspace{2pt}
$B[\mbox{start}+i] = 1$, \hspace{7pt}
$\Str'[\mbox{start}+X_s+i-X[i]] = \Str[\mbox{start}+i]$
\Else \hspace{1pt}
$B[\mbox{start}+i] = 0$, \hspace{7pt}
$\Str'[\mbox{start}+X[i]] = \Str[\mbox{start}+i]$
\EndIf
\EndFor
\vspace{-2pt}
\If {$X_s > 0$} 
$A'[2*j] = (\mbox{start}, X_s, 2*\mbox{id} +1, r/2)$
\Comment{Left child}
\EndIf
\vspace{-2pt}
\If {$(\mbox{len} - X_s) > 0$}
$A'[2*j+1] = (\mbox{start}+X_s, \mbox{len}-X_s, 2*\mbox{id}+2, r/2)$
\Comment{Right child}
\EndIf
\EndIf
\EndFor
\vspace{-2pt}
\State Filter out empty $A'$ entries and store into $A$
\State swap($\Str,\Str'$)
\EndFor
\vspace{-2pt}
\State \textbf{return} Nodes
\EndProcedure
\end{algorithmic}
\vspace{-2pt}
\caption{levelWT: Level-by-level parallel algorithm for wavelet tree construction.} \label{alg:levelWT}
\vspace{-5pt}
\end{figure}

Initially, the array $A$ contains just the root node, with a starting
index of 0, length of $n$ (it represents all elements), node ID of 0,
and a range of $2^{\lceil \log \sigma \rceil}$ (Line 2). The array
$A'$ is used as the output array for the level. The algorithm
proceeds one level at a time for $\lceil \log \sigma\rceil$ levels
(Line 3). The bitmaps on each level are determined by the $l+1$'st
highest bit in the symbols, so we use a mask to determine the sign of
this bit in the symbols (Line 4). 
%% Each node determines its
%% offset into the bitmap $B$ using a
%% prefix sum, with the offsets stored in the array $O$ (Lines 5--7). 
The
algorithm then loops through all the nodes on the current level in
parallel (Lines 5--21). For each node, it sets its bitmap pointer and
length in the Nodes array (Line 7). If the alphabet range of the node
is 2 or less, then it has no children (Line 8). The algorithm then
just loops over the node's symbols in $\Str$ in parallel, and sets
each bit in the bitmap according to the sign of the symbol's $l+1$'st
highest bit (Lines 9--10).\footnote{\scriptsize The actual code requires bit
  arithmetic to access the appropriate bit in the word, which we
  omit for simplicity.} Otherwise, the symbols in $\Str$ are
rearranged (and stored into $\Str'$) so that they are in the correct
order on the next level. To do this in parallel, we first count the
number of symbols that go to the left child ($l+1$'st highest bit
is 0), and the offset of each such symbol using a prefix sum (Lines
12--14). With the prefix sum array $X$ and result $X_s$, the symbols
that go to the right child can be computed as well using the formula
$X_s+i-X[i]$ (the number of symbols with an $l+1$'st highest bit of 0
is $X_s$, so this is the number of symbols to offset, and then the
number of symbols with an $l+1$'st highest bit of 1 up to index $i$ is
$i-X[i]$).
%% (this is similar to rank queries on binary sequences). 
In Lines 15--17, the bits in the bitmap and positions in $\Str'$ are
set in parallel.  Children nodes are placed into $A'$ on Lines 18--19
if the number of symbols represented for the child is greater than 0.
The children's node IDs are computed as twice the current node ID plus
one for the left child and twice the current node ID plus two for the
right child in order to give all nodes unique IDs. The starting point
in the bitmap of the next level is the same as the current starting
point for the left child, and is the current starting point plus the
number of elements on the left ($X_s$) for the right child. The length
is stored from the computation before. The range of each child is half
of the current range.
%% If the current range is
%% $r$ then the range of the left child is $2^{\lceil \log r \rceil - 1}$
%% and range of the right child is $r - 2^{\lceil \log r \rceil -1}$.
After each level, the non-empty entries of $A'$ are filtered out
and stored into $A$ (Line 20). The roles of
$\Str$ and $\Str'$ are swapped for the next level (Line 21).

We note that setting the bits in $B$ (Lines 10, 16 and 17) must be
done atomically since multiple processors may write to the same word
concurrently. This can be implemented using a loop with a
compare-and-swap until successful. An optimization we use is to only
perform atomic writes if a word is shared between two nodes
(there can be at most two shared words per node, as the bits for each
node are contiguous in $B$), and for the remaining words we
parallelize at the granularity of a word. Inside each word, the
updates are done sequentially. This allows us to use
regular writes for all but at most two words per node.

We also note that we are able to stop early when the range of a node
is 2 or less since the construction of the previous level provides
this information. The algorithm of~\cite{Fuentes14} (and also our next
algorithm) processes all levels independently, so it is not easy to
stop early.
%% We observe that the high-level
%% structure of levelWT is similar to a stable radix sort that processes
%% bits one at a time from the most significant bit to the least
%% significant bit.%, but with some additional operations and bookkeeping.

We now analyze the complexity of levelWT. For each level, there is a
total of $O(n)$ work performed, since each symbol is processed a
constant number of times. The prefix sum on Lines 14 and the filter on
Line 20 require linear work and $O(\log n)$ depth per level. The
parallel for-loops on Lines 9--10, 13 and 15--17 require linear work
and $O(1)$ depth per level.  There are $O(\log \sigma)$ levels so the
total work is $O(n\log\sigma)$ and depth is $O(\log
n\log\sigma)$. Since $\sigma \le n$, the depth is polylogarithmic in
$n$.  We obtain the following theorem:
\begin{theorem}
levelWT requires
$O(n\log\sigma)$ work and $O(\log n\log\sigma)$ depth.
\end{theorem}

We now describe our second wavelet tree construction
algorithm, which constructs all levels of the wavelet tree in
parallel. We refer to this algorithm as
\emph{sortWT}. Since preceding levels cannot provide information to
later levels, we must do independent computation per level to obtain
the correct ordering of the sequence for the level.  We will make use
of the observation of Fuentes-Sepulveda et al.~\cite{Fuentes14} that
the node at which a symbol $s$ is represented at level $l$ of the
wavelet tree ($l=0$ for the root) is encoded in the top $l$
bits. For level $l$, our algorithm sorts $\Str$ using the top $l$ bits
as the key, which gives the correct ordering of the sequence for the
level. We note that the sort must be stable since the relative
ordering of nodes with the same top $l$ bits must be preserved in the
wavelet tree. 
%% Since the keys are in a bounded range ($\sigma \le n$),
%% we can use parallel stable integer sorting.

%% Using a stable parallel integer sort requires $O(n)$
%% work and $O(\sigma^{\epsilon}+\log n)$ depth for ($O <
%% \epsilon < 1)$~\cite{RR89,JaJa92}.
%% The depth is polylogarithmic only when $\sigma$ is polylogarithmic. We
%% can also use a stable parallel integer sort requiring $O(n\log\log n)$
%% work and $O(\log n)$ depth, but it is not work-efficient. Therefore the
%% total work is $O(W_{sort}\log\sigma)$ and depth is $O(S_{sort})$ where
%% $W_{sort}$ and $S_{sort}$ are the work and depth, respectively, of the
%% stable sorting routine that we use.  Since $O(n)$ space is required
%% per level, and since auxiliary arrays cannot be reused among levels,
%% the total space complexity is $O(n\log\sigma)$.

\begin{figure}[!t]\scriptsize
\vspace{-3pt}
\begin{algorithmic}[1]
\Procedure{sortWT}{$\Str$, $n$, $\sigma$}
\State Nodes = $\{\}$,\hspace{7pt} $L = \lceil \log \sigma \rceil$
\ParFor {$l = 0$ to $L-1$}
\State mask = $2^{L-(l+1)}$,\hspace{7pt} $B =$ bitmap of length $n$ \Comment{Mask and bitmap for this level}
\If {$l = 0$} \Comment{No sorting required for first level}
\ParFor {$i=0$ to $n-1$}
{$\{$ {\bf if} {$(\Str[i] \ \&\ \mbox{mask} \neq 0)$} 
$B[i] = 1$ 
{\bf else}
$B[i] = 0$ $\}$}
\EndFor
\vspace{-2pt}
\State Nodes[0].bitmap = 0, \hspace{7pt} Nodes[0].len = $n$
\vspace{-1pt}
\Else 
\State \hspace{1pt} $\Str'$ = $\Str$ stably sorted by top $l$ bits
\ParFor {$i=0$ to $n-1$}
{$\{$ {\bf if} {$(\Str'[i] \ \&\ \mbox{mask} \neq 0)$} 
$B[i] = 1$ 
{\bf else}
$B[i] = 0$ $\}$}

\EndFor
\vspace{-2pt}
\State $O = $ indices $i$ such that $(\Str'[i] \gg L - l) \neq
(\Str'[i-1] \gg L-l)$ \Comment{Using a filter}
\ParFor{$i=0$ to $|O|-1$}
\State id = $2^l-1 + (\Str'[O[i]] \gg L-l)$,\hspace{7pt} Nodes$[\mbox{id}]$.bitmap = $O[i]$,\hspace{7pt} Nodes$[\mbox{id}]$.len = $O[i+1]-O[i]$
\EndFor
\EndIf
\EndFor
\vspace{-2pt}
\State \textbf{return} Nodes
\EndProcedure
\end{algorithmic}
\vspace{-2pt}
\caption{sortWT: Sorting-based parallel algorithm for wavelet tree construction.} \label{alg:fp-WT}
\vspace{-5pt}
\end{figure}

The pseudo-code for sortWT is shown in Figure~\ref{alg:fp-WT}. As in
levelWT, we use a mask and a bitmap $B$ for each level (Line 4).  For
the first level ($l=0$), no sorting of $\Str$ is required, so we
simply fill the bitmap according to the highest bit of each symbol
(Lines 5--6).  For each subsequent level, we first stably sort $\Str$
by the top $l$ bits to obtain the symbols in the correct order, and
store it in $\Str'$ (Line 9). The bitmap is filled according to the
$l+1$'st highest bit of each symbol (Line 10). To compute the length
of each node's bitmap we use a filter to find all the indices where
the symbol's top $l$ bits differ from the previous symbol's top $l$
bits in $\Str'$ (Line 11). These mark the bitmaps of each node since
$\Str'$ is sorted by the top $l$ bits. The length of each bitmap can
be computed by the difference in indices. On lines 12--13, we set the
bitmap pointers and lengths for the nodes on the current level. The
IDs of the nodes start at $2^l-1$, since there are up to $2^l-1$ nodes
in previous levels, and each node ID is offset by the top $l$ bits of
the symbols that it represents, as this determines the node's position
in the level.  As in levelWT, updates to the bitmaps are done in
parallel at word granularity. % so atomic updates are not required.
%% Again, the updates to the bitmaps
%% need to be atomic, but we can apply the optimization discussed for the
%% levelWT algorithm. The optimization will not be done on a per-node
%% basis but for the entire level.

We now discuss the algorithm's complexity. Let $W_{sort}(n)$ and
$D_{sort}(n)$ be the work and depth, respectively, of the stable
sort on Line 9. The filter on Line 11
requires $O(n)$ work and $O(\log n)$ depth. The parallel for-loops on
Lines 6, 10 and 12--13 require $O(n)$ work and $O(1)$ depth. The
overall work is $O(W_{sort}(n)\log\sigma)$ and since all levels can be
computed in parallel, the overall depth is $O(D_{sort}(n)+\log
n)$. This gives the following theorem:
\begin{theorem}
sortWT requires $O(W_{sort}(n)\log\sigma)$ work and
$O(D_{sort}(n)+\log n)$ depth.
\end{theorem}

Using $\frac{1}{\epsilon}$ rounds of linear-work parallel stable
integer sorting~\cite{RR89} ($W_{sort}(n) = O(\frac{n}\epsilon)$ and
$D_{sort}(n) = O(\frac{1}\epsilon(\sigma^{\epsilon}+\log n))$ for some
constant $0 < \epsilon < 1$), we obtain a work bound of
$O(\frac{n}\epsilon\log\sigma)$ and depth bound of
$O(\frac{1}\epsilon(\sigma^{\epsilon}+\log n))$, which is
sub-linear. For $\sigma = O(\log^c n)$ for any constant $c$, this
gives $O(\log n)$ depth (by setting $\epsilon$ appropriately). We can
alternatively use a stable integer sorting algorithm with
$W_{sort}(n)= O(n\log\log n)$ and $D_{sort} = O(\frac{\log n}{\log\log n})$
(either using randomization~\cite{Raman1990} or using super-linear
space~\cite{Bhatt1991}) to obtain an overall work of $O(n\log\log
n\log\sigma)$ and depth of $O(\log n)$ for any alphabet size.

%% We have
%% the following theorem:
%% \begin{theorem}
%% msortWT requires $O(n\log\sigma)$ work and $O(\log
%% n\log\sigma)$ depth.
%% \end{theorem}

%% Theoretically we can improve the depth by constructing in parallel
%% chunks of $O(\log\log n)$ levels (the $i'th$ level in the chunk
%% sorting by $i$ bits), and saving the result of the last level in the
%% chunk to be used for the next chunk. This will involve keys of at most
%% $O(\log\log n)$ bits, or keys in a range up to $O(\log n)$, so the
%% integer sort can then still be done in $O(\log n)$ depth. This improves
%% the overall depth to $O(\log n\log\sigma/\min(\log\sigma,\log\log n))$,
%% or equivalently $O(\log n\max(1,\log\sigma/\log\log n))$.

\subsec{Space usage} The input and output of the algorithms is
$O(n\log\sigma)$ bits.  levelWT requires two auxiliary arrays for the
prefix sum on each level (that can be reused per level), which takes
$O(n\log n)$ bits. For sortWT, since all levels are processed in
parallel, and each level requires $O(n\log n)$ bits for the integer
sort, the total space usage is $O(n\log n\log\sigma)$ bits.  

Due to the high space usage and hence memory footprint of sortWT, we
found that processing the levels one-by-one gives better performance
in practice, as will be discussed in Section~\ref{sec:experiments}
(although this increases the depth by a factor of $O(\log\sigma)$). We
call this modified version \emph{msortWT}.  On each level, msortWT
sorts the sequence from the previous level.  Since
the levels are processed one-by-one, msortWT has a space usage of
$O(n\log n)$ bits.

\section{Experiments}\label{sec:experiments}
\subsec{Implementations} We compare our implementations of levelWT,
sortWT and msortWT with existing parallel implementations as well as a
sequential implementation.  The implementations all use the levelwise
representation of the bitmaps, where one bitmap of length $n$ is
stored per level, and nodes have pointers into the bitmaps. sortWT and
msortWT use linear-work parallel stable integer sorting.
%% For
%% levelWT, atomic writes to the bitmaps are only performed when a word
%% on the bitmap is shared by another node.  
We use the parallel prefix sum, filter and integer sorting
routines from the Problem Based Benchmark Suite~\cite{SBFG}. 
We compare with the implementations of
Fuentes-Sepulveda et al.~\cite{Fuentes14}, one which computes each level of
the wavelet tree independently in parallel (\emph{FEFS}), and one 
%% We observed that the speedup of
%% the code was limited for large alphabets as it performed a number of
%% memory allocations proportional to the number of nodes in the wavelet
%% tree in parallel, which causes high contention. 
which computes a partial wavelet tree for each thread, and then merges them
together (\emph{FEFS2}). Both implementations require the alphabet size
to be a power of 2, so we only report times for the inputs with such
an alphabet size.
%% , but
%% unfortunately, we were unable to get it to run on most input files.
We
implement a sequential version of wavelet tree construction
(\emph{serialWT}), and found its performance to be competitive with the times
of the sequential algorithm reported in~\cite{Grossi2011}. We also
tried the serial implementation in SDSL~\cite{Gog2014} for
constructing a balanced wavelet tree, but found it to be slower than
serialWT on our inputs.  However, the SDSL implementation is more
space-efficient, and sometimes faster on the Burrows-Wheeler
transformed inputs. A comparison with SDSL is
presented in %the full version of the paper~\cite{full}.
Section~\ref{sec:bwt} of the Appendix.

\subsec{Experimental Setup} We run our experiments on a 40-core (with
two-way hyper-threading) machine with $4\times 2.4\mbox{GHz}$ Intel
10-core E7-8870 Xeon processors (with a 1066MHz bus and 30MB L3
cache), and 256\mbox{GB} of main memory.  The parallel codes use Cilk
Plus, and we compile our code
with the \texttt{g++} compiler version 4.8.0 (which supports Cilk
Plus) with the \texttt{-O2} flag. The times reported are based on a
median of 3 trials.

We use a variety of real-world and artificial sequences. The
real-world sequences include strings from {\footnotesize
  \url{http://people.unipmn.it/manzini/lightweight/corpus/}}, XML code
from Wikipedia (\emph{wikisamp}), protein data from
{\footnotesize \url{http://pizzachili.dcc.uchile.cl/texts/protein/}}
(\emph{proteins}), the human genome from {\footnotesize
  \url{http://webhome.cs.uvic.ca/~thomo/HG18.fasta.tar.gz}}
(\emph{HG18}), and a document array of text collections
(\emph{trec8}). The artificial inputs (\emph{rand}), parameterized
by $\sigma$, are generated by drawing each symbol uniformly at random
from the range $[0,\ldots,\sigma-1]$. 
%% We use these sequences to
%% experiment with large alphabets.
%% We did not artificially vary the
%% entropy of the sequences, as the entropy does not significantly affect
%% construction time~\cite{Grossi2011}.  
The lengths and alphabet sizes of the inputs are listed in
Table~\ref{table:times}. 
%% For the alphabets with range at most 256, we
%% used 8 bits per symbol to represent the input sequence, and otherwise
%% we used 32 bits per symbol. We modified the FEFS code
%% to do so as well.

Due to the various choices for rank/select structures, each of which
has different space/time trade-offs, we do not include their
construction times in the wavelet tree construction time.
%%  (we note that
%% the time for rank/select structure construction is small compared to
%% wavelet tree construction).  
We modified the FEFS and FEFS2 code
accordingly. The times for our code include generating the
parent/child pointers for the nodes, although these could be
removed using techniques from~\cite{MakinenN07,ClaudeN08}.  FEFS and
FEFS2 do not generate these pointers.

\begin{table}[!t]
\vspace{-3pt}
\scriptsize
\centering
\begin{tabular}{c|c|c|c|c|c|c|c|c|c|c|c|c|c}\\
Text & $n$ & $\sigma$ & serialWT & levelWT & levelWT & sortWT & sortWT & msortWT & msortWT & FEFS & FEFS & FEFS2 & FEFS2\\
 &  &  & ($T_1$) & ($T_1$) & ($T_{40}$) & ($T_1$) & ($T_{40}$) & ($T_1$) & ($T_{40}$) & ($T_1$) & ($T_{40}$) & ($T_1$) & ($T_{40}$) \\
\hline\hline
chr22 & $3.35\cdot 10^7$ & 4 & 0.486 & 0.611 & 0.018 & 0.786 & 0.046 & 0.768 & 0.029 & 1.03 & 0.53 & 0.98 & 0.1\\
etext99 & $1.05 \cdot 10^8$& 146 & 4.12 & 6.99 & 0.28 & 10.5 & 0.393 & 9.79 & 0.364 & -- & -- & -- & --\\
HG18 & $2.83\cdot 10^9$& 4 & 35.5 & 45.5 & 1.32 & 57.7 & 1.88 & 56.9 & 1.67 & 76.6 & 39.1 & 72.2 & 2.55\\
howto & $3.94\cdot 10^7$& 197 & 1.65 & 2.69 & 0.105 & 4.07 & 0.161 & 3.86 & 0.144 & -- & -- & -- & --\\
jdk13c & $6.97\cdot 10^7$& 113 & 2.51 & 3.9 & 0.159 & 6.13 & 0.234 & 5.63 & 0.22 & -- & -- & -- & -- \\
proteins & $1.18\cdot 10^9$& 27 & 31.3 & 52.2 & 1.82 & 75.3 & 2.59 & 71.3 & 2.28 & -- & -- & -- & --\\
rctail96 & $1.15\cdot 10^8$& 93 & 3.54 & 5.88 & 0.231 & 10.2 & 0.373 & 9.39 & 0.34 & -- & -- & -- & --\\
rfc & $1.16\cdot 10^8$& 120 & 3.8 & 6.51 & 0.261 & 10.1 & 0.37 & 9.28 & 0.348 & -- & -- & -- & --\\
sprot34 & $1.1\cdot 10^8$& 66 & 3.69 & 6.26 & 0.248 & 9.48 & 0.36 & 8.8 & 0.328 & -- & -- & -- & --\\
trec8 & $2.43\cdot 10^8$ & 528155 & 33.3 & 50.4 & 2.08 & 138 & 5.5 & 104 & 4.55 & -- & -- & -- & --\\

w3c2 & $1.04\cdot 10^8$& 256 & 3.82 & 6.66 & 0.275 & 10.6 & 0.388 & 9.78 & 0.357 & 11.1 & 2.0 & 10.6 & 0.51\\
wikisamp & $10^8$ & 204 & 3.52 & 6.16 & 0.264 & 9.78 & 0.374 & 9.08 & 0.349 & -- & -- & -- & --\\
rand-$2^{8}$ & $10^8$ & $2^8$ & 5.76 & 8.58 & 0.36 & 14.3 & 0.652 & 12.1 & 0.533 & 12.4 & 1.71 & 12.3 & 0.5\\
rand-$2^{10}$ & $10^8$ & $2^{10}$ & 6.88 & 11 & 0.456 & 19 & 0.857 & 15.9 & 0.708 & 15.0 & 1.71 & 15.3 & 0.58\\
rand-$2^{12}$ & $10^8$ & $2^{12}$ & 8.32 & 12.4 & 0.525 & 24.5 & 1.11 & 20.4 & 0.922 & 18.7 & 1.78 & 17.4 & 0.67\\
%rand-$2^{14}$ & $10^8$ & $2^{14}$ & 9.81 & 14.2 & 0.595 & 30.1 & 1.35 & 24.9 & 1.13 & &\\
rand-$2^{16}$ & $10^8$ & $2^{16}$ & 11.2 & 16.4 & 0.655 & 36 & 1.59 & 29.5 & 1.34 & 34.4 & 3.26 & 32.4 & 1.28 \\

rand-$2^{20}$ & $10^8$ & $2^{20}$ & 14 & 20.4 & 0.772 & 49.5 & 2.19 & 40.6 & 1.85 & 65.7 & 7.14 & 64.8 & 3.94\\
%random-$2^{24}$ & & $2^{24}$\\
\end{tabular}

%\vspace{-2pt}
\caption{Comparison of running times (seconds) of
    wavelet tree construction algorithms on a
    40-core machine with hyper-threading. $T_{40}$ is the time using
    40 cores (80 hyper-threads) and $T_1$ is the time using a single
    thread.}\label{table:times}
\vspace{-3pt}
\end{table}

\subsec{Results} Table~\ref{table:times} shows the single-thread
($T_1$) and 40-core with two-way hyper-threading ($T_{40}$) running
times on the inputs for the various implementations.  
%% The parallel
%% times reported are the best time for all thread counts between 1 and
%% 80.
%% For FEFS2, we only report
%% numbers for the inputs on which the code ran successfully on.
Our results show that levelWT is faster than sortWT and msortWT both
sequentially and in parallel. This is because sortWT and msortWT use
sorting, which has a larger overhead. msortWT is slightly faster than
sortWT due to its smaller memory footprint.
%%  and also because for each
%% level it already has a partially sorted sequence so there is less data
%% movement from the sort. 
Compared to serialWT, levelWT is 1.2--1.8x
slower on a single thread, and 13--27x faster on 40 cores
with hyper-threading. The self-relative speedup of levelWT ranges from
23 to 35. On 40 cores, sortWT and msortWT are 6--19x
and 7--22x faster than serialWT, respectively.  sortWT and
msortWT achieve self-relative speedups of 17--31 and 22--34,
respectively.

FEFS2 always outperforms FEFS on 40 cores with two-way hyper-threading
because FEFS splits the work among only $\log\sigma$ threads, which is
less than 80 on all our inputs ($\sigma < 2^{80}$), whereas FEFS2
splits the work among all available threads in its first step. The
second (merging) step of FEFS2, however, only makes use of
$\log\sigma$ threads, but this is a smaller fraction of the total
time.  On 40 cores with hyper-threading, our best implementation
(levelWT) outperforms FEFS2 by a factor of 1.3--5.6x, and FEFS by much
more.  Compared to msortWT, FEFS2 is faster in some cases and slower
in others.

%% This is because FEFS and FEFS2 only have $O(\log\sigma)$
%% parallelism whereas the parallelism for our algorithms is much higher
%% ($O(n/\log n)$ for levelWT, $O(n\log\sigma/(\sigma^\epsilon+\log n))$
%% for sortWT and $O(n/(\sigma^\epsilon+\log n))$ for msortWT). The
%% parallelism in our algorithms is much higher than the number of cores
%% available, and thus we achieve very good speedup.

\begin{figure*}[!t]
\centering
\vspace{-3pt}
\subfigure{
\includegraphics[width=0.4\linewidth]{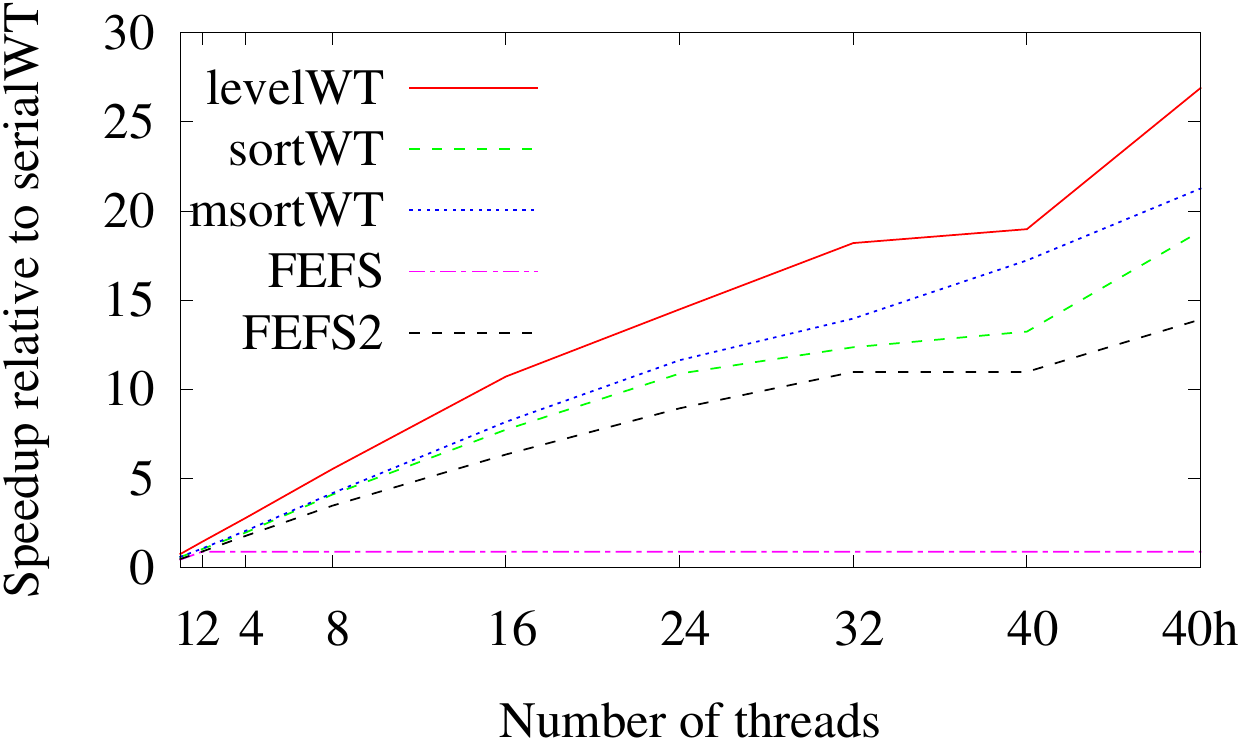}
\label{fig:breakdown-TC-merge}
}
%% \subfigure{
%% \includegraphics[width=0.31\linewidth]{figs/pdfs/w3c2_speedup.pdf}
%% \label{fig:breakdown-TC-hash}
%% }
\subfigure{
\includegraphics[width=0.4\linewidth]{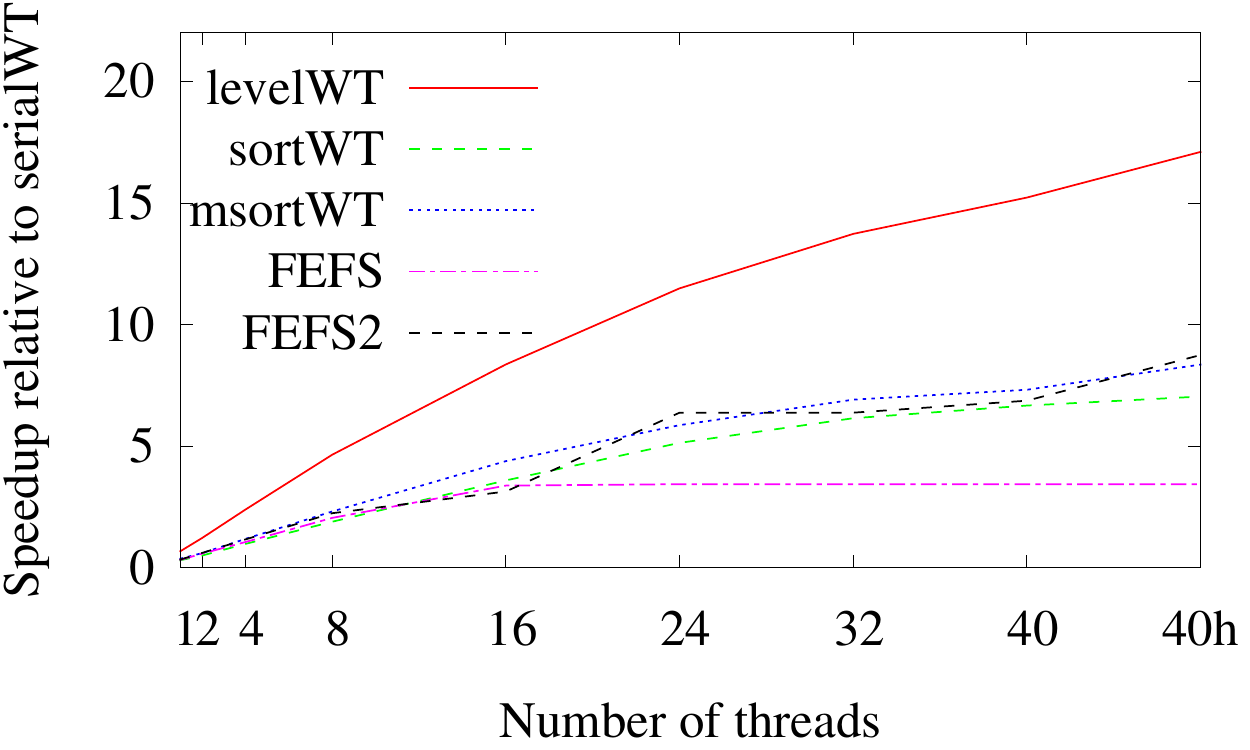}
\label{fig:breakdown-TC-approx}
}
\vspace{-4pt}
\caption{Speedup of implementations relative to serialWT for HG18
  ({\bf left}) and rand-$2^{16}$ ({\bf right}). (40h) is 40
  cores with hyper-threading.}
\label{fig:speedup}
\vspace{-2pt}
\end{figure*}

In Figure~\ref{fig:speedup}, we plot the speedup of the parallel
implementations relative to serialWT as a function of the number of
threads for HG18 and rand-$2^{16}$. For HG18, our implementations and
FEFS2 scale well up to 80 hyper-threads. FEFS only scales up to 2 threads
due to the small alphabet size.
%%  however FEFS and FEFS2 only scales up to 2
%% threads and 4 threads, respectively due to the small alphabet. 
For rand-$2^{16}$, our implementations again exhibit good scalability.
FEFS scales up to 16 threads as there are 16 levels in the tree, while
FEFS2 scales to more threads. FEFS2 is competitive with msortWT, but
slower than levelWT.

%% For
%% FEFS, scalability is limited since it does $2^{16}$ memory allocations
%% in parallel, which leads to high contention, and so it scales only up
%% to 4 threads. FEFS2 is competitive with levelWT for up to 8 threads
%% but with higher numbers of threads it does not scale well due to the
%% linear-depth merge step.

\begin{figure*}[!t]
\centering
\vspace{-3pt}
\subfigure{
\includegraphics[width=0.4\linewidth]{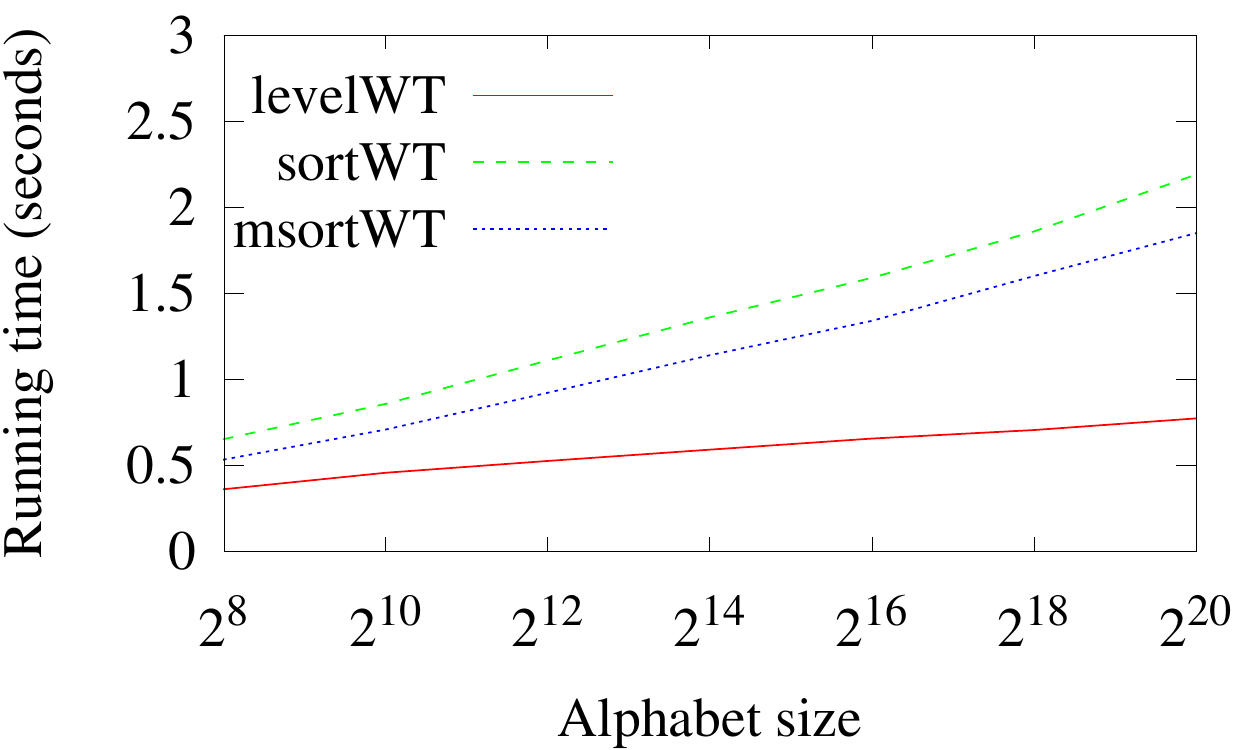}
\label{fig:sigma}
}
%\hspace{20pt}
\subfigure{
\includegraphics[width=0.4\linewidth]{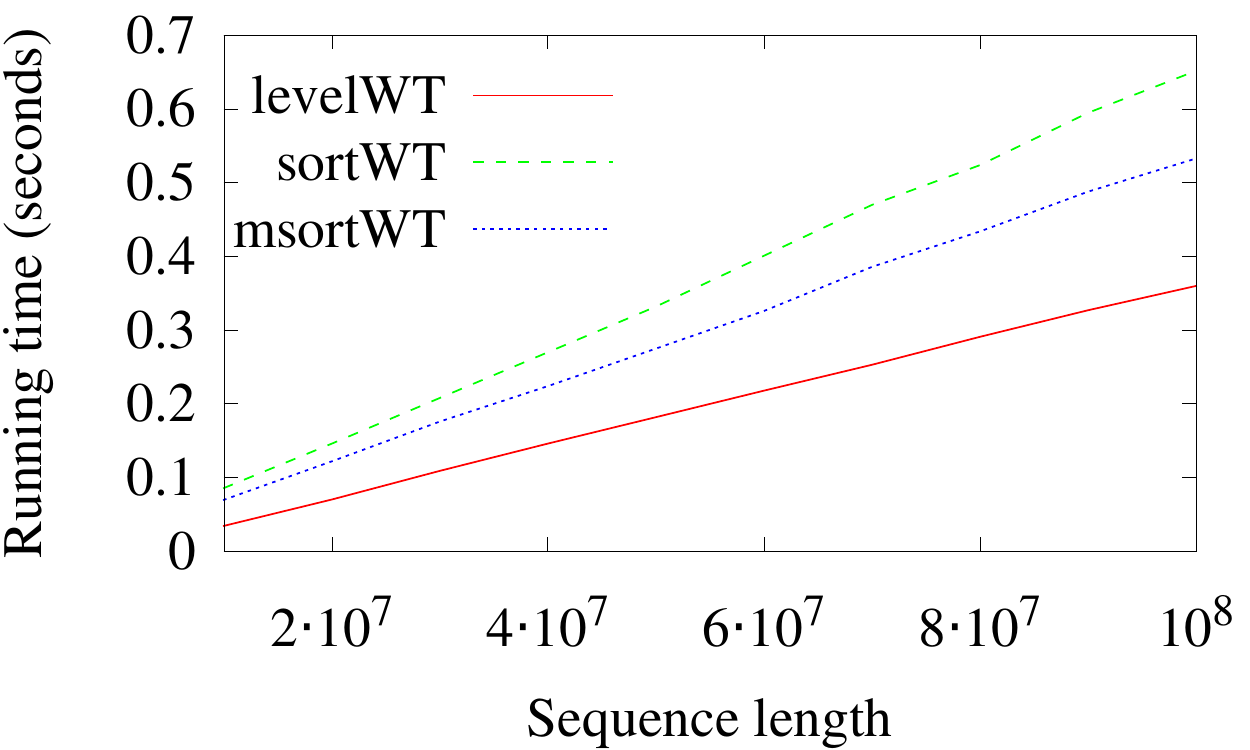}
\label{fig:n}
}
\vspace{-4pt}
\caption{40-core (with hyper-threading) running times vs. $\sigma$ ({\bf left}, $x$-axis in log-scale) and vs. $n$ ({\bf right}) on random sequences ($\sigma = 2^8$).}
\label{fig:scale}
\vspace{-4pt}
\end{figure*}

In Figure~\ref{fig:scale} (left), we plot the 40-core parallel running
time of our three implementations as a function of the alphabet size
for random sequences of length $10^8$. We see that for fixed $n$, the
running times increase nearly linearly with $O(\log\sigma)$, which is
expected since the total work is $O(n\log\sigma)$. In
Figure~\ref{fig:scale} (right), we plot the running time of the
implementations as a function of $n$ for $\sigma = 2^8$ on random
sequences, and see that the times increase linearly with $n$ as
expected. Our algorithms exhibit similar
speedups on 40 cores as we vary $\sigma$ or $n$, since the core count
is much lower than the available parallelism of the algorithms.

In summary, our experiments show that our three parallel algorithms
for wavelet tree construction scale well with the number of threads,
input length and alphabet size. levelWT outperforms sortWT and msortWT
as it does not have the overheads of sorting, and achieves good
speedup over serialWT. Overall, our fastest algorithm outperforms FEFS
and FEFS2.
%% although for many inputs FEFS2 is not that much slower because it uses
%% all available threads in its first step.

%% with a modest number of threads as they exhibit much more parallelism
%% (poly-logarithmic instead of linear depth).

%% Describe optimization with atomic OR. Times for construction with and
%% without tree pointers.

\section{Parallel Construction of Rank/Select Structures}\label{sec:rank-select}
Wavelet trees make use of succinct rank/select structures which
support constant work rank and select queries on binary sequences.
%Constructing these structures sequentially requires $O(n)$ work.  
In this section, we describe how to construct these structures in
parallel using $O(n)$ work and $O(\log n)$ depth for a binary sequence
of length $n$.
%% parallelize the construction of such
%% a structure, which is necessary to achieve our depth
%% bounds for wavelet tree construction.
We note that \emph{sequential} construction of
rank/select structures in $o(n)$ work have been described
in~\cite{BabenkoGKS14,Munro15}, however parallel construction in linear work suffices
for our purposes.

%% Note that the algorithms
%% of~\cite{Fuentes14} do not need a parallel construction of rank/select
%% structures since they require $O(n)$ depth. 
%% We use the rank structure of Jacobson~\cite{Jacobson1988} and select
%% structure of Clark~\cite{Clark1996}.
%% Later results~\cite{Brodnik1999,Pagh2001,Raman2002} reduce the
%% space usage of the rank/select structures even further, and their
%% constructions can be parallelized using similar techniques.

The rank structure of Jacobson~\cite{Jacobson1988} stores the rank of
every $\log^2n$'th bit in a first-level directory, and the rank of
every $\log n$'th bit in each of the ranges in a second-level
directory. Rank queries in each range of size $\log n$ can be answered
by at most two table look-ups, where the table stores the rank of all
bit-strings of length up to $\frac{\log n}2$.  The first- and
second-level directories of can be constructed by converting the
bit-string to a length $n$ array of 0's and 1's and computing a prefix
sum on the array in $O(n)$ work and $O(\log n)$ depth. Entries in the
second-level directory require $\log\log n$ bits each, and for space
efficiency, they need to be packed into words. This can be done by
processing groups of $O(\frac{\log n}{\log\log n})$ entries (the number
that fits in a word) in parallel, and packing each group into a word
sequentially. There are $O(\frac{n}{\log n})$ entries, and word
operations take $O(1)$ work and depth per entry, so this process takes
$O(n)$ work and $O(\log n)$ depth.
%% since there are $O(\frac{n}{\log n})$
%% entries and the word length is $O(\log n)$.  
The look-up table can be
constructed in $o(n)$ work and $O(\log n)$ depth, as we can compute the
number of 1's in bit-strings of size $O(\log n)$ in $O(\log n)$ work
and depth, and there are $O(2^{\frac{\log n}2}\log n) = O(\sqrt{n}\log
n)$ such bit-strings.

Clark's select structure~\cite{Clark1996} stores the position of every
$\log n\log\log n$'th 1 bit in a first-level directory. Then for
each range $r$ between the positions, if $r \ge \log^2n(\log\log n)^2$
the $\log n\log\log n$ answers in the range are stored
directly. Otherwise the position of every $\log r\log\log n$'th 1
bit is stored in a second-level directory. The sub-ranges $r'$ in the
second-level directory are again considered, and if $r'\ge \log r'\log
r(\log\log n)^2$, then all answers in the range are stored
directly. Otherwise, a look-up table is constructed for all
bit-strings of length less than $r'$.
%% , which can be
%% shown to be of size $O(\log \log n)$.
To parallelize the construction, we first convert the bit-string to an
array of 0's and 1's, and compute the positions of all the 1 bits
using a prefix sum and filter in $O(n)$ work and $O(\log n)$
depth. This allows all of the ranges to be processed in
parallel. Constructing each second-level directory again uses prefix
sum and filter. Over all directories, this sums to $O(n)$ work and
$O(\log n)$ depth. The look-up table can be constructed in $o(n)$ work
and $O(1)$ depth, similar to the rank structure. Packing entries into
words can also be done within the complexity bounds.

We note that there have been more practical variants of rank/select structures
(see, e.g.,~\cite{ZhouAK13,GogP2013} and references
therein) that have a similar high-level structure.
% to the ones described above. 
%% The most recent ones~\cite{ZhouAK13,GogP2013} follow a
%% similar structure to the constructions described above.
%% , and can also
%% be parallelized in $O(n)$ work and $O(\log n)$ depth.

%~\cite{Navarro12,ZhouAK13,OkanoharaS07,Gonzalez05,ClaudeN08,Vigna2008,Ladra2012}. 
%% The recent solution of Zhou et al.~\cite{ZhouAK13} can be
%% parallelized. Their rank structure follows the framework described
%% above with the following minor differences: (1) it uses three levels
%% of directories, (2) stores the second and third levels in interleaved
%% form, and (3) uses the \texttt{popcnt} (population count) instruction
%% to compute the number of 1's in the small blocks. The construction is
%% parallelizable as before, and interleaving can be done by adjusting
%% the directory indices. Their select structure only uses one directory
%% and is parallelizable as before.

\section{Extensions}
The \emph{Huffman-shaped wavelet tree}, where each node is placed at a
level proportional to the length of its Huffman code, was introduced
to improve compression and average query
performance~\cite{Foschini2006}. To construct it in parallel, we first
compute the Huffman tree and prefix codes in $O(\sigma+n)$ work and
$O(\sigma+\log n)$ depth using the algorithm of~\cite{Edwards2014}.
The construction then follows the strategy of levelWT.  To decide how
to set the bitmaps in the internal nodes and rearrange $\Str$, we must
know the side of the tree that each symbol is located on. We map each
symbol to an integer corresponding to the location of its leaf in an
in-order traversal of the Huffman tree, which can be done in parallel
using an Euler tour algorithm in $O(\sigma)$ work and $O(\log\sigma)$
depth~\cite{JaJa92}. At each internal node, the symbols to the left
and to the right are in consecutive ranges, and we store the highest
mapped integer of nodes in its left sub-tree, which can be done during
the Euler tour computation.  Then the decision of how to set the
bitmap and where to place the symbol in $\Str'$ can be made with a
single comparison with the mapped integers. The rest of the
computation follows the logic of levelWT.  For a tree of height $h$,
the overall work (including Huffman encoding) is $O(nh)$ as linear
work is done per level.
%% as the sum of
%% bitmap sizes (proportional to the number of times symbols are processed) is $O(n\log\sigma)$~\cite{Foschini2006}.
The overall depth is $O(\sigma+h\log n)$, as each level of the tree
takes $O(\log n)$ depth. 
%Similar ideas can be used to construct binary wavelet trees of other shapes.
% as long as the tree shape and labels on the nodes can be computed efficiently.

%Hu-Tucker wavelet tree~\cite{BarbayN09}

%% Ferragina et al.~\cite{Ferragina2007} generalize the
%% succinct rank/select structures for binary sequences to alphabets of
%% size $o(\log n/\log\log n)$, while still providing constant time
%% queries. To support general alphabets, they 

Ferragina et al.~\cite{Ferragina2007} describe the 
\emph{multiary wavelet tree} where each node has up to $d$ children
for some value $d$, and stores sequences of symbols in the range
$[0,\ldots,d-1]$. The height of the tree is $O(\log_d\sigma)$. 
%% For $h
%% = \log n/\log\log n$, the structure can answer queries in $O(\log
%% h/\log\log n)$ time (using a rank/select structure for larger
%% alphabets).  
%% We can modify the version of sortWT described in the
%% ``Relating sortWT to levelWT'' sub-section of Section~\ref{sec:alg} to
%% construct the multiary wavelet tree.  
We describe the parallel construction for the case where $d =
O(\log^{\epsilon}n)$ for $\epsilon<1/3$, and $d$ is a power of
two.\footnote{\scriptsize The requirement $\epsilon<1/3$ is necessary
  for the analysis of the rank/select structure~\cite{BabenkoGKS14}
  that we parallelize.}
%$\log h$ is an integer.
%% , though it can be modified
%% for the general case.  
We modify sortWT to process the levels one-by-one and save the sorted
sequence $\Str'$ for the next level.  On level $l$, $\Str'$ is already
sorted by the top $(l-1)\log d$ bits, so we only need to sort the next
$\log d$ highest bits within each of the sub-sequences sharing the
same top $(l-1)\log d$ bits. The sub-sequence boundaries can be
identified with a filter, and for each sub-sequence we apply a
stable integer sort using the $\log d$ appropriate bits for the level as the
key.  We substitute the bitmap $B$ with a sequence of $\log d$-bit
entries, and in parallel each entry is set according to the value of
the appropriate $\log d$ bits of each symbol. There will be up to
$d^l$ nodes on level $l$, and the offset to the Nodes array is
$d^l-1$.
%% \footnote{\scriptsize Note that for $d=2$, this description
%%   gives a variant of sortWT for constructing the standard wavelet tree
%%   with a high-level structure similar to that of levelWT.}
%% We now discuss a modified version of sortWT (\emph{msortWT}) which has
%% $O(\log n\log\sigma)$ depth. Instead of processing all levels in
%% parallel, we process the levels one by one. We 
%% When $h = O(\log^cn)$ for a constant $c$, which is true in applications
%% of multiary wavelet trees, 
Since $d = O(\log^{\epsilon} n)$, the integer sort on each level requires
$O(n)$ work and $O(\log n)$ depth, giving an overall work of
$O(n\log_d\sigma)$ and depth of $O(\log
n\log_d\sigma)$.
%% The depth can be improved to $O(\log
%% n\log_h\sigma/\min(\log_h\sigma,\log\log n))$ by grouping the
%% levels. 
%% One can use the original sortWT algorithm to improve the depth to
%%  $O(\log n)$ when $\sigma = O(\log^c n)$.
%, or $O(\log n)$ depth and
%% $O(n\log\log n\log_h\sigma)$ work for larger alphabets.  
We also parallelize the construction of rank/select structures on sequences with larger
alphabets, used in the multiary wavelet tree.  
The description is 
in Section~\ref{sec:generalized} of the Appendix.
%the full version of the paper~\cite{full}.
% leave for further work.

The \emph{wavelet matrix}~\cite{Claude12} is a variant of the wavelet
tree where on level $l$, all symbols with a 0 as their $l$'th
highest bit are represented on the left side of the level's bitmap
and all symbols with a 1 as their $l$'th highest bit are
represented on the right. Each level stores the number of 0's on
the level. The bitmap is filled based on the $l+1$'st highest bit of
the symbols.
%% It is shown that this representation still allows for
%% logarithmic time queries~\cite{Claude12} while being faster in
%% practice. 
To construct the wavelet matrix, we proceed level-by-level and stably
reorder $\Str$ based on the $l$'th highest bit of the symbols using
standard operations involving prefix sum (similar to levelWT) in
$O(n)$ work and $O(\log n)$ depth, which also gives the number of 0's
on the level.  The bitmap for the level is then filled in parallel in
$O(n)$ work and $O(1)$ depth.  This gives an algorithm with
$O(n\log\sigma)$ work and $O(\log n\log\sigma)$ depth.  We can
alternatively use a strategy similar to sortWT, but using the reverse
of the top $l$ bits as the key when sorting.
%% Since there are no dependencies among levels, we can compute all
%% levels in parallel, reducing the depth to $O(\log n)$, but increasing
%% the space usage.

%% Grossi and Ottaviano describe the wavelet trie, an extension of the
%% wavelet tree for maintaining a dynamic sequence over a dynamic
%% alphabet~\cite{Grossi2012}. The parallelization of its construction
%% and of performing batched updates is an interesting direction for
%% future work.

%\input{alg3}
%\input{conclusion}

%% \bibliographystyle{abbrv}
%% {\footnotesize 
%% \bibliography{ref}}
%\input{todo}
%\appendix

\begin{spacing}{0.85}

{
\setlength{\bibitemsep}{0.2pt}
\setlength{\bibhang}{0em}
\setlength{\biblabelsep}{0.5em}
\renewcommand{\bibfont}{\footnotesize}
\printbibliography
}
 \end{spacing}

 \appendix
 \section{Appendix}
\subsection{Performance on Burrows-Wheeler transformed inputs}\label{sec:bwt}
For certain applications (see~\cite{Navarro2012,Makris12}), the
wavelet tree is constructed on the Burrows-Wheeler transform of the
sequence.  In Table~\ref{table:bwt-times}, we report the running times
of the algorithms on the Burrows-Wheeler transform of the real-world
texts.  For sequential times, we also report the wavelet tree
implementation from SDSL~\cite{Gog2014}, which performs well on
sequences with many sequences of repeated characters. It uses an
optimization that groups the writes into the bitmap for repeated
characters into a single write. This gives an advantage for the
Burrows-Wheeler transformed texts, in which there are often many
sequences of repeated characters. In contrast to serialWT, the SDSL
implementation generates the wavelet tree structure first before
updating the bitmaps.
%% This requires an additional pass over the input, however
%% saves some computation on determining what bits to write.  
In the column labeled \emph{SDSL} in Table~\ref{table:bwt-times}, we
report the SDSL times using the balanced tree option and without the
construction of the rank/select structures.  For our implementations,
we use an optimization where consecutive 1 bits in a word are written
together instead of individually.\footnote{\scriptsize We thank
  Matthias Petri for suggesting this optimization.} This is similar to
SDSL but it finds consecutive 1 bits on each level instead of
consecutive repeated characters in the original sequence. The
optimization slightly improves the running time for the
Burrows-Wheeler transformed texts, but makes negligible difference on
the original texts.  For some inputs, SDSL performs slightly faster
than serialWT, though it is slower than or performs about the same as
serialWT on other inputs.

%% although it was slightly slower on
%% the original text due to the additional computational overhead as the
%% chunks of ``1'' bits are not as large.

\setlength{\tabcolsep}{2.3pt}

\begin{table}[!t]
%\vspace{-3pt}
\scriptsize
\centering
\begin{tabular}{c|c|c|c|c|c|c|c|c|c|c|c|c|c|c}\\
Text & $n$ & $\sigma$ & serialWT & SDSL & levelWT & levelWT & sortWT & sortWT & msortWT & msortWT & FEFS & FEFS & FEFS2 & FEFS2\\
 &  &  & ($T_1$) & ($T_1$) & ($T_1$) & ($T_{40}$) & ($T_1$) & ($T_{40}$) & ($T_1$) & ($T_{40}$) & ($T_1$) & ($T_{40}$) & ($T_1$) & ($T_{40}$) \\
\hline\hline
chr22 & $3.35\cdot 10^7$ & 4 & 0.48 & 0.86 & 0.59 &  0.017 & 0.762 & 0.044 & 0.741 & 0.029 & 1.0 & 0.51 & 0.95 & 0.09\\
etext99 & $1.05 \cdot 10^8$& 146 & 2.83 & 2.9 & 5.79 & 0.266 & 9.47 & 0.377 & 9.04 & 0.343 & -- & -- & -- & --\\
HG18 & $2.83\cdot 10^9$& 4 & 31 & 63.2 & 36.7 & 1.17 & 43.3 & 1.6 & 41.4 & 1.42 & 71.9 & 37.9 & 67.2 & 2.53\\
howto & $3.94\cdot 10^7$& 197 & 1.2 & 1.2 & 2.14 & 0.1 & 3.58 & 0.157 & 3.42 & 0.14 & -- & -- & -- & -- \\
jdk13c & $6.97\cdot 10^7$& 113 & 1.51 & 1.13 & 2.88 & 0.152 & 5.23 & 0.223 & 4.9 & 0.207 & -- & -- & -- & --\\
proteins & $1.18\cdot 10^9$& 27 & 26.1 & 30.2 & 43.9 & 1.69 & 65.1 & 2.34 & 61 &2.11 & -- & -- & -- & -- \\
rctail96 & $1.15\cdot 10^8$& 93 & 2.11 & 2.31 & 4.61 & 0.221 & 8.83 & 0.35 & 8.28 & 0.327 & -- & -- & -- & --\\
rfc & $1.16\cdot 10^8$& 120 & 2.53 & 2.73 & 5.26 & 0.251 & 8.95 & 0.355 & 8.52 & 0.335 & -- & -- & -- & --\\
sprot34 & $1.1\cdot 10^8$& 66 & 2.55 & 2.6 & 5.19 & 0.24 & 8.53 & 0.343 & 8.14 & 0.31 & -- & -- & -- & --\\
%trec8 & $2.43\cdot 10^8$ & 528155 & 33.3 &\\
w3c2 & $1.04\cdot 10^8$& 256 & 2.3 & 1.85 & 5.13 & 0.263 & 9.76 & 0.375 & 8.6 & 0.349 & 9.44 & 1.94 & 8.53 & 0.47\\
wikisamp & $10^8$ & 204 & 2.22 & 1.91 & 5.06 & 0.253 & 8.67 & 0.353 & 8.17 & 0.325 & -- & -- & -- & --\\
\end{tabular}

%\vspace{-2pt}
\caption{Running times (seconds) of wavelet tree construction
  algorithms on a 40-core machine with hyper-threading \emph{on the
    Burrows-Wheeler transform of the text}. $T_{40}$ is the time using
  40 cores (80 hyper-threads) and $T_1$ is the time using a single
  thread.}\label{table:bwt-times}
%\vspace{-3pt}
\end{table}

Overall, the times are faster on the Burrows-Wheeler transformed texts
than on the original texts. This is because the bits of the same value
are grouped into larger chunks in the bitmaps, and hence there are
fewer words that actually need to be updated (words with all 0 bits do
not need to be updated after initialization). Furthermore, the
optimization of writing consecutive 1 bits together has more of a
benefit in this setting. The relative performance among our parallel
algorithms remains the same, with levelWT being the fastest, followed
by msortWT and then sortWT. Compared to the faster of serialWT and
SDSL for each input, levelWT is 1.7--2.8 times slower on a single
thread and 7--28 times faster on 40 cores with hyper-threading.  The
self-relative speedups of levelWT, sortWT and msortWT are 18--35,
17--28 and 24--29, respectively.  Our implementations are always
faster than FEFS and FEFS2 in parallel, though FEFS2 gets good speedup.

\subsection{Parallel Construction of Generalized Rank/Select Structures}\label{sec:generalized}
A generalized rank/select structure supports rank and select queries
on sequences with larger alphabets in $O(1)$ work. The rank query
$rank_c(\Str,i)$ returns the number of symbols less than or equal to $c$
in $\Str$ from positions $0$ to $i$ (in contrast to the definition in
Section~\ref{sec:prelims}).

We construct in parallel the rank/select structure described in the
Appendix of~\cite{BabenkoGKS14}, which works for an alphabet size
of $\sigma = O(\log^{\epsilon} n)$ for $\epsilon<1/3$. We only discuss
its construction, and refer the reader to~\cite{BabenkoGKS14} for
discussions on the space usage of the resulting structure. Again, the
sequential construction process described in~\cite{BabenkoGKS14} takes
sub-linear work, but we describe a parallel algorithm that takes
linear work, which suffices for our purposes.

\subsec{Rank}
The rank structure stores the $\sigma$ ranks (one per character) of
every $\sigma\log^2n$'th symbol in a first-level directory, and the
$\sigma$ ranks of every $\log n/(3\log\sigma)$'th
symbol in a second-level directory. Queries on blocks of up to length
$\log n/(3\log\sigma)$ can be answered in constant
work using table lookup. The lookup table has at most
$O(\sigma^{2+\log n/(3\log\sigma)}) = o(n)$ entries,
so can be created in $o(n)$ work and $O(\log n)$ depth.  

Computing ranks for the second-level directory entries again uses
table lookup.  Sequentially, to compute the ranks of an entry we take
the ranks of the previous entry and update it with the block of $\log
n/(3\log\sigma)$ symbols between them. The results of updating the
ranks of all possible second-level directory entry along with all
possible blocks of $\log n/(3\log\sigma)$ symbols can be pre-computed
in a table of size $o(n)$ again in $o(n)$ work and $O(\log n)$ depth.

We now need to parallelize the computation of the ranks of the
second-level directory entries.  We will do this by applying a prefix
sum on the entries with the combining operator defined by a lookup
table. Since a prefix sum computation uses a reduction
tree~\cite{JaJa92} and will combine entries separated by more than
$\log n/(3\log\sigma)$ symbols, we cannot directly apply the table
lookup scheme described above. We fix this as follows.
%% Recall that a prefix sum is implemented using a contraction
%% phase that combines all the values into an overall ``sum'', followed
%% by an expansion phase which propagates the partial sums to the
%% respective offsets~\cite{Vishkin10}.  
During the prefix sum, a combined entry $e$ will represent more than
one original entry, and we store the ranks of the first and last entry
that it represents (call these $e_F$ and $e_L$). Without loss of
generality, assume $e$ combines with an entry $e'$ before it in the
input. We then use the lookup table described previously to update
$e_F$ given $e'_L$ and the block of $\log n/(3\log\sigma)$ symbols between
them. All ranks of $e_L$ will then be increased by the same amount as
in $e_F$.  When combining $e$ and its previous entry $e'$, we keep
$e'_F$ and $e_L$. Each combine operation takes constant work, so the
prefix sum takes $O(n)$ work and $O(\log n)$ depth.

The ranks of a first-level directory entry can be computed by adding
the ranks of the previous first-level entry and the ranks of the last
second-level entry associated that first-level entry. This can be
parallelized with prefix sum, where the combining operator adds all
$\sigma$ ranks in parallel.  In the reduction tree, each combined
entry $e$ keeps track of the sum of ranks of first-level entries it
represents ($e_s$) as well as last second-level entry $e_t$ of its
last first-level entry. When combining $e$ with an entry $e'$ before
it in the input, we use we create a new combined entry $e''$ which
stores the sum of $e_s$ and $e'_t$ as its sum, and $e_t$ as its last
second-level entry.  Recall that there are only $n/(\sigma\log^2n)$
first-level entries. Thus, this prefix sum takes $o(n)$ work and
$O(\log n)$ depth.

Therefore, the construction of the rank structure takes linear work
and $O(\log n)$ depth.  Packing the entries into words to meet the
space requirements can be done within the same bounds.

\subsec{Select} For select, a separate structure is stored for each
character in the alphabet along with an array of size $O(\sigma)$ with
pointers to the structures. The structure for character $c$ stores the
location of every $\sigma\log^2n$'th occurrence of $c$ in a
first-level directory. Then for each range $r$ between the locations,
if $r \ge \sigma^2\log^4n$ then the answers in the range are directly
stored. Otherwise we store the occurrence of every $\sigma(\log\log
n)^2$'th of $c$ in the range in a second-level directory. The
sub-ranges in the second-level directory are then considered. If
the length of a sub-range is at least $\sigma^3(\log\log n)^4$, then
the answers are stored directly. Otherwise the range is small enough
such that we can use a lookup table to find the $k$'th occurrence of
$c$ in the range in $O(1)$ work. The lookup table can be constructed
in $O(n^d)$ work for $d<1$ and $O(\log n)$ depth.

We compute the entries for all $\sigma$ structures together.  To
compute the first-level directory entries, we first split the input
into chunks of size $\sigma$. Each chunk sequentially computes the
number of occurrences of each character by reading characters
one-by-one and updating the count associated with the character read
as in~\cite{BabenkoGKS14}. This requires $O(n)$ work and $O(\sigma)$
depth. Then we perform a prefix sum on the results of the $n/\sigma$
chunks, so that at the end of each chunk we know how many occurrences
of each character there are up to its location. This requires a total
of $O(n)$ work and $O(\log n)$ depth. With this information, we can
compute which chunks the first-level directory entries lie in for each
character, and store them in a packed array using prefix sum in $O(n)$
work and $O(\log n)$ depth. Note that for each character there can be
at most one first-level entry per chunk since the chunk size is less
than $\sigma\log^2n$.  In parallel for all chunks that contain a
first-level entry, we go into the chunk and find it
sequentially. Since the chunk is of length $\sigma$, this takes $O(n)$
work and $O(\sigma)$ depth.  For each character $c$ we can now compute
the ranges $r$ between first-level entries. We mark the chunks
corresponding to the ranges $r \ge \sigma^2\log^4n$, as well as store
the starting and ending point in the chunk. This takes $O(n)$ work and
$O(1)$ depth.  Using the initial prefix sum result, we can allocate
the appropriate amount of space to store the entries that need to be
stored directly inside these ranges, as well as compute appropriate
offsets into the shared space among different chunks.  Now we read the
chunks in parallel and if the character read is inside a range (which
can be checked in $O(1)$ work) we output its location to the
appropriate offset in the corresponding select structure. This takes
$O(n)$ work and $O(\sigma)$ depth.  

We then similarly mark the chunks corresponding to the ranges $r <
\sigma^2\log^4n$, and output every $\sigma(\log\log)^2$'th occurrence
of a character. This gives the sub-ranges $r'$ for each character $c$
and we can again mark the chunks based on whether $r'$ is at least
$\sigma^3(\log\log n)^4$, and perform the appropriate operations which
can be done in $O(n)$ work and $O(\sigma)$ depth using the same ideas
as above.

The entire construction of the select structure takes $O(n)$ work and
$O(\sigma+\log n)$ depth. Since $\sigma = o(\log n)$, the depth
simplifies to $O(\log n)$.  Again, packing the entries into words to
satisfy the space requirements can be done in the same bounds.

%% Lastly, each
%% entry in the second step takes the result of the prefix sum and
%% propagates the updates to the remaining entries of its group, which
%% can be done with table lookup on a similar table as described above. The first and last steps take $O(n\log\alpha/\log n)$ work and $O(\log n)$ depth as all groups are processed in parallel and each group is of size $O(\log n)$. We must be careful in analyzing the second step because the entries being combined in the prefix sum are separated by more than $\frac{1}{3}\frac{\log n}{\log\sigma}$ symbols and require more than $O(1)$ table lookups. 
%% Prefix sum is implemented using a reduction tree of $O(\log n)$ height~\cite{Vishkin10}, and each level of recursion will perform 
%% $3n\log\alpha/\log^2 n$

%% the number of table lookups is no longer $O(1)$ as the two entries being combined could be separated by many more symbols. 
%% However because the recursion tree depth is logarithmic and each level of recursion performs $O(n

\end{document}